\documentclass[a4paper,11pt]{article}
\usepackage{amsmath,amsfonts,amssymb,times,mathptmx,graphicx,amsthm,lineno}
\usepackage{placeins}
\usepackage[T1]{fontenc}
\usepackage[latin1]{inputenc}
\title{\bf Probability Distributions of Positioning Errors for Some Forms of
Center-of-Gravity Algorithms.
 }
\author{Gregorio Landi$^a$\thanks{Corresponding
author. Gregorio.Landi@fi.infn.it}~,   Giovanni E. Landi$^b$\\
\\
\llap{$^a$} Dipartimento di Fisica e Astronomia,
Universita' di Firenze and INFN\\
Largo E. Fermi 2 (Arcetri) 50125, Firenze, Italy\\
\\
\llap{$^b$} ArchonVR S.a.g.l.,\\
Via Cisieri 3,
6900 Lugano, Switzerland.\\ \\
{ April 18, 2020}}
\date{ }
\begin{document}
\maketitle 
\begin{abstract}
The center of gravity is a widespread algorithm for position reconstruction
in particle physics.
For track fitting, its standard use is always accompanied by an easy  guess for the
probability distribution of the positioning errors. This is an incorrect
assumption that degrades the results of the fit. The explicit error forms show evident
Cauchy-(Agnesi) tails that render problematic the use of variance minimizations.
Here, we report the probability distributions for some combinations of
random variables, impossible to find in literature, but essential for track fitting:
$x={\xi}/{(\xi+\eta)}$, $y={(\xi-\eta)}/[2{(\xi+\eta)}]$,
$w=\xi/\eta$, $x=\theta(x_3-x_1) (-x_3)/(x_3+x_2) +\theta(x_1-x_3)x_1/(x_1+x_2)$
and $x=(x_1-x_3)/(x_1+x_2+x_3)$.
The first three are directly connected to each other and are partial forms of
the two-strip center of gravity. The fourth is the complete two-strip center
of gravity. For its very complex form, it allows only approximate expressions
of the probability.
The last expression is a simplified form of the three-strip center of gravity. General
integral forms are obtained for all of them. Detailed analytical
expressions are calculated assuming $\xi$, $\eta$,
$x_1$, $x_2$ and $x_3$ independent random variables with
Gaussian probability distributions (the standard assumption for the strip noise).
\end{abstract}

\newpage


\tableofcontents

\pagenumbering{arabic} \oddsidemargin 0cm  \evensidemargin 0cm


\section{Introduction }

Very complex probability density functions (PDFs) are fundamental
tools to obtain the resolution improvements in the simulations
of refs.~\cite{landi05,landi06,landi07}.
These PDFs were appositely developed to describe the statistical
properties of the positioning algorithms for signals of
minimum ionizing particles (MIPs) in silicon micro-strip detectors.
Their construction was motivated by the observation of
the impossibility, for a single PDF, to produce the
distributions of the simulated data (scatter-plots).
The observed scatter-plots were those of ref.~\cite{landi03}.
They illustrated samples of center of gravity (COG)
calculated with MIP signals reported as a function of the particle
impact point ($\varepsilon$). To explore the importance of additional
pieces of information, those simulations were used
to produce seven very rough approximations of the $\varepsilon$-PDFs
for a fixed interval of COG values. These rough PDFs were used
in a maximum likelihood search for parameters of reconstructed straight
tracks of MIPs. The evident improvements of the parameter
distributions, compared to those of the standard fits (least squares),
convinced us about the importance of these additional
pieces of information. For an extensive study of these hints,
more accurate forms of PDFs were essential, as illustrated
in refs.~\cite{landi05,landi06,landi07}.
As consequence of those results, refs.~\cite{landi08,landi09}
demonstrate that the standard fitting
methods are non-optimal just for the neglect of
the hit differences. In fact, we proved that the
standard fits have parameter variances always
greater than the parameter variances of
fits accounting for the hit
properties (variances).

The aim of this work is to complete the methods
employed in the previous publications giving the explicit
expressions for the used PDFs. The calculated PDFs
refer to the center of gravity (COG) algorithm.
The COG algorithm is an easy and
efficient positioning algorithm of large use in
particle physics.
The generic COG definition is $\sum_jE_jX_j/\sum_jE_j$,
where $E_j$ are the signals of a cluster inserted in
the COG and $X_j$ their positions. Different selections
of the signals inserted in the COG expression generate
a set of positioning algorithms with different analytical
and statistical properties.
Our special attention is directed  to the two strip COG (COG$_2$)
for its minimal noise. The COG$_2$ is computed with the signals of
the leading strip (seed) and the maximum of the two contiguous strips.
Its PDF has  a typical gap, the explanation of this feature and an example
of it is reported in ref.~\cite{landi01}.
It is just the reproduction of this typical feature
that renders very complicated the calculation and the form of the
PDF for the COG$_2$.
Nevertheless, the COG$_2$ PDF was extensively used in
the simulations of ref.~\cite{landi05,landi06} with very
large improvements of the track parameters.
Even if our attention is focalized on COG$_2$, also other
COG PDFs will be illustrated, few of them of large use.
However, the COG PDFs are only a part of the
complications in track fitting,
the other part is the insertion of a dependence
from the particle impact point ($\varepsilon$).
Completed with the impact point, the
PDF becomes able to take into account the signal-to-noise ratio
of each strip and to  correct the COG systematic errors
(ref.~\cite{landi01}).
The insertion of the $\varepsilon$-dependence requires
the exploration and filtering of special types
of random processes and the availability of large
sets of homogeneous data as delineated in ref.~\cite{landi05}.
Further details about the handling of these types of random
processes will be discussed elsewhere.
In section 2, the convenience to go beyond the least squares
method is illustrated and the simplest forms of COG PDFs are reported.
Section 3 and 4 are devoted to the complete COG$_2$ PDF and the
PDF for the three strip COG. Two appendices, one with a derivation
of a simple COG PDF from the cumulative probability distribution
and the other with a better (and longer) approximation of the COG$_2$ PDF,
complete this (partial) illustration of the COG PDFs.
These results are obtained with an extended use of
MATHEMATICA~\cite{MATHEMATICA} and verified in many
realistic cases with MATLAB~\cite{matlab} simulations.

\section{Definition of the problem}

It is easy to observe, (as in ref.~\cite{landi03}),
the non-uniformity of the point distributions in  scatter
plots of COG simulations.
In ref.~\cite{landi05}, these differences are better
illustrated with the definition of an effective
variance for each hit and with distributions of
samples of these values. These distributions
substantially differ from a horizontal line,
the obvious result of a single PDF and its single
variance. Thus, the hypothesis of a
single PDF, for the positioning errors, must be ruled out
in favor of more realistic assumptions.
In fact, it easy to grasp the effects, on a fit, given
by the possibility to distinguish
good or excellent hits from average or worst hits.
The corrections of the hit properties, due to
the differences of detector technologies
along the lines of ref.~\cite{hartmann}, are small steps
in the right direction but absolutely insufficient.
Experimental indications about differences of the hit properties
are reported in ref.~\cite{CMS_13}. However, the Landau
distribution of the charge released by a MIP is another
well known experimental result that adds further differences
to the hit properties.
The maximum likelihood method allows the use of all the
information contained in the data, and it is able to
give the drastic improvements of the track parameters even in
presence of outliers,  as discussed in
ref.~\cite{landi05}. This ability to handle outliers
is a consequence of the tails of PDFs.
Another consequence is due to the different quality
of hit PDFs, as discussed above, two goods (or excellent) hits suffice for a
good (or excellent) straight track fit, and
the probability of good (or excellent) hits grows with the
number of hits (detecting layers) per track~\cite{landi07,landi08,landi09}.
Thus, the pool of the track parameters is enriched
at this rate. Instead, the standard least squares
grows as the square root
of the number of detecting layers. A very slow
growth respect to the linear one, with a waste
of tracker resolution and  an
increases of tracker complexity.

In spite of the proofs of the
maximum likelihood as the best fitting
method, intrinsic difficulties
limits its use. For the very complex trackers of the LHC experiments,
its full machinery is probably beyond the allowed computer resources.
Even if the schematic approximations of ref.~\cite{landi05}
reduces the maximum likelihood method to a weighted least squares,
the computing of the effective variances for each hit
requires large CPU-time. However, with negligible imprecisions,
very fast look-up tables can be constructed for the hit-effective
variances. Or the lucky model of ref.~\cite{landi07} can be an easy
substitute with a small loss of resolution. It must be remembered that
the schematic approximation and the
lucky model are ineffective on the outliers.

In any case, the maximum likelihood method, in its full extension or the
schematic approximation, requires the use of analytic expressions
of the PDFs with the general functional forms $P(\varepsilon,x_{g_n})$.
Where, $x_{g_n}$ is a generic COG with n-strips and
$\varepsilon$ is the MIP impact point. The conditional probabilities
$P(\varepsilon|x_{g_n})$ and $P(x_{g_n}|\varepsilon)$ are connected to
the marginal probabilities $P(x_{g_n})$ and $ P(\varepsilon)$ as usual:

\begin{equation*}
    P(\varepsilon|x_{g_n})P(x_{g_n})=P(\varepsilon,x_{g_n})=
    P(x_{g_n}|\varepsilon)P(\varepsilon)\,.
\end{equation*}
The constant probability of the impact point $\varepsilon$ is assured by:
\begin{equation}\label{eq:equation_1a}
    P(\varepsilon)=\int_{-\infty}^{+\infty}P(\varepsilon,x_{g_n})\, \mathrm{d}x_{g_n}=1\,,
\end{equation}
and it is consistent with the assumption of uniformity we used in
ref.~\cite{landi03}, and its normalization on a strip.
We will see that this condition is granted by
the normalization of the PDFs.

The Kolmogorov axioms~\cite{gnedenko} attributes to the cumulative
probability distribution a the fundamental role to calculate the PDF.
The cumulative probability distribution for a continuous case
is given by integrations on the appropriate portion
of the plane, or the space, as the geometry
of the problem requires.
Differentiating the cumulative probability distribution gives
the PDF.
This method becomes extremely long with complicated algorithms.
However, our first approach was modeled on
the ratio of two random variables as described in
ref.~\cite{gnedenko}, and we followed the
method with the cumulative distributions
for all our PDFs, from the simplest
to the most complex one. This very long set of
integrals is too boring to be reported in a paper,
and this is the principal reason for the delay of
this report.
Here,  we will utilize a different approach,
very direct and flexible, with use of Dirac $\delta$-functions
and Heaviside-$\theta$ functions, operating directly on PDFs.
An assay was given in ref.~\cite{landi06}.
This method is a variance of the Fermi golden rule $\# 1$
that is extensively used for the cross-section calculations
(or diffusion probabilities), and it recovers the results of
our geometric approaches. To underline the consistency
with the geometric approach, the first part of the COG$_2$
PDF will be obtained with the cumulative probability distribution
in Appendix A.

It will be assumed that the random signals are the charges
released on the strips by the hitting particle.
The signals are corrupted by
additive random noises, of Gaussian PDFs, produced by
the rest of the acquisition system. The data
are at their final elaboration procedure (calibration, pedestal,
common noise suppression etc.) and are ready to be used in
a positioning algorithm of any type. The stream of primary charges, released by a
MIP in the detector, diffuse on a cluster of strips.
The charges collected by a strip are correlated with those
collected by the cluster. The distributions of the charges
in the cluster depend, among other parameters as particle direction and total
released charges, from the MIP impact point. Hence, the
$\varepsilon$ dependence of $P(\varepsilon,x_{g_n})$ is contained in
the strip signal $a_i$. Here, we will consider the signals $a_i$
as parameters and the PDF will be expressed in the form $P(\{a_i\},x_{g_n})$.
The variable $x_{g_n}$ will be abandoned for a more simpler
$x$. The strip size is always taken to be one, and it is the length scale
of the system. For our definitions, the parameters $a_i$ can be expressed
in any dimensional units consistent with those of the noise$\sigma_i$
(we use directly the ADC counts).  The variable $x$ turns out to be a
pure number as the PDFs.

Each strip has its own random additive noise uncorrelated
with that of any other strip. In absence of MIP signal, the strip noise is
well reproduced with a Gaussian PDF. Thus, the PDFs for the
signal plus noise of the strip $i$ become:

\begin{equation}\label{eq:equation_2}
    P_i(z)=\frac{\exp[-\frac{(z-a_i)^2}{2\sigma_i^2}]}{\sqrt{2\pi}\,\sigma_i}\ \ \ \ i=1,2,\cdots
\end{equation}
The Gaussian mean values $\{a_i\}$ are the (noiseless) charges
collected by the strips and are positive numbers (we assume to
handle signals from real particles).
The parameters $\{\sigma_i\}$ are the standard deviations of the
additive zero-average Gaussian noise.

\subsection{Probability for the ratio $x=\xi/(\xi+\eta)$}

The first  explored PDF is the distribution of the
random values of $x$ with $x=\xi/(\xi+\eta)$. This
expression has the structure of a COG with the origin of
the reference system in the
center of the strip $\#2$ ($\eta$ random variable) and another signal on the
right strip $\# 1$ ($\xi$ random variable).
This form of COG is the right part of the full COG$_2$ algorithm.
For its limitation to only two random variable $\{\xi,\eta\}$,
it is a first step toward more complex PDFs. The derivation of the PDF
for $P_{xg_2R}(x)$ with the cumulative distribution is
illustrated in Appendix A. However, this PDF
can be rapidly obtained with the method discussed
in the following.

\begin{equation}\label{eq:equation_1}
    P_{xg_2R}(x)=\frac{1}{x^2}\int_{-\infty}^{+\infty}
    P_1(z)P_2\big(\frac{1-x}{x}z\big)\,|z\,| \mathrm{d} z
\end{equation}
The heavy tail of a Cauchy-like distribution is evident.
Equation~\ref{eq:equation_1} shows a $1/x^2$  behavior for $x\rightarrow\infty$,
and the factor $(1-x)/x$ goes to $-1$. In this limit, the integral is
convergent and different from zero.
The singularity for $x=0$ does not exist (because the integral goes to zero),
and it can be removed with the coordinate transformation
$z/x=\zeta$. But, it is preferable to save the $1/x^2$ factor to
remember the divergence of the variance for $P_{xg_2R}(x)$.
The Gaussian integral is analytic for any $x$ and $a_i$,  and has the form:
\begin{equation}\label{eq:equation_5}
    \begin{aligned}
    P_{xg_2R}(x)=&\Big\{\frac{a_2(1-x)\sigma_1^2+a_1x\,\sigma_2^2}{\sqrt{2\pi}[(1-x)^2\sigma_1^2+x^2\sigma_2^2]^{3/2}}\exp\big[-
    (\frac{a_1}{a_1+a_2}-x)^2\frac{(a_1+a_2)^2}{2(\sigma_1^2(1-x)^2+x^2\sigma_2^2)}\big]\,\\
    &\mathrm{erf}\big[\frac{a_2(1-x)\sigma_1^2+a_1x\,\sigma_2^2}
    {\sqrt{2}\sigma_1\sigma_2\sqrt{(1-x)^2\sigma_1^2+x^2\sigma_2^2}}\big]\Big\}+
    \exp\big[-\frac{a_1^2}{2\sigma_1^2}-\frac{a_2^2}{2\sigma_2^2}\big]
    \frac{\sigma_1\sigma_2}{\pi[(1-x)^2\sigma_1^2+x^2\sigma_2^2]}\,.\\
    \end{aligned}
\end{equation}
The form of the $P_{xg_2R}(x)$ shows some aspects that will be
found often in the following. It is easy to recognize, in eq.~\ref{eq:equation_5},
part of the PDF reported in ref.~\cite{landi06}.
Equation~\ref{eq:equation_5}
has a maximum for $x\approx a_1/(a_1+a_2)$. This point is
the noiseless COG for this variable combination and,
on average, tends to eliminate the COG systematic error
of ref.~\cite{landi01}.
Around the maximum, $P_{xg_2R}(x)$ looks similar
to a Gaussian. However, the exponential becomes very different
from a Gaussian for large $x$, where it
goes to a non-zero constant. The modulating term of the
maximum is connected to the signal to noise ratio
of the two strips. The positivity of the PDF is
granted by a term $A\, \mathrm{erf}(A)$ that for a
not too small $A$ converges rapidly to $|A|$.
Around zero, $A\, \mathrm{erf}(A)$ is a
continuous differentiable function and it
differs from $|A|$ essentially for the cusp at $A=0$ of $|A|$.
The range of the differences respect to $|A|$ are of the order
of $\sigma_1$ (or some weighted average with $\sigma_2$).
This range is expected to be negligible, if the detection
algorithm works well and discards almost all the fake hits (with $a_i=0$).
Thus very often we will substitute $A\, \mathrm{erf}(A)$ with $|A|$.\newline
The last term will be called Cauchy-like term, it is very
similar, but not identical, to a Cauchy PDF.
This term survives even for $a_1=a_2=0$ and could be a
probability of fake hits. It assures the strict positivity
of the PDF. For $a_i\neq 0$ is heavily suppressed by the
negative exponents, quadratic
in the strip signal-to-noise ratio.

The validity of this PDF is limited to one side of the
COG$_2$ algorithm. The track reconstruction requires a rigid
connection to the local reference system, naturally
centered in the seed-strip center. Thus, it is important
to conserve a difference from the left strip, the central
strip, and the right strip. The track impact point $\varepsilon$ can be located
even outside the seed strip.

Another PDF, that composes the complete COG$_2$ PDF, contains the
random variable $\beta$, the signal collected by
strip $\# 3$ positioned to the left of the
strip $\# 2$. This PDF will be indicated as
$P_{xg_2L}(x)$. As for $P_{xg_2R}(x)$, it will be
assumed that the strip $\# 2$ is the the seed of
the strip cluster. As always, the origin of the reference
system is the center of the strip $\#2$.
Now, we have for $x$ the combination of random
variables $-\beta/(\beta+\eta)$.
The function $P_{xg_2L}(x)$ is obtained from
eq.~\ref{eq:equation_5} with the
substitution $a_1\rightarrow a_3$, $\sigma_1\rightarrow\sigma_3$
and $x\rightarrow -x$. We report here $P_{xg_2L}(x)$,
often in the following, terms of this type will be
indicated with the substitutions needed for their construction.

\begin{equation}\label{eq:equation_6}
    \begin{aligned}
    P_{xg_2L}(x)=&\Big\{\frac{a_2(1+x)\sigma_3^2-a_3x\sigma_2^2}
    {\sqrt{2\pi}[(1+x)^2\sigma_3^2+x^2\sigma_2^2]^{3/2}}\exp\big[-
    (\frac{a_3}{a_3+a_2}+x)^2\frac{(a_3+a_2)^2}{2(\sigma_3^2(1+x)^2+x^2\sigma_2^2)}\big]\,\\
    &\mathrm{erf}[\frac{(1+x)a_2\sigma_3^2-a_3 \,x\sigma_2^2}
    {\sqrt{2}\sigma_3\sigma_2\sqrt{(1+x)^2\sigma_3^2+x^2\sigma_2^2}}]\Big\}+
    \exp[-\frac{a_3^2}{2\sigma_3^2}-\frac{a_2^2}{2\sigma_2^2}]
    \frac{\sigma_3\sigma_2}{\pi[(1+x)^2\sigma_3^2+x^2\sigma_2^2]}\,.\\
    \end{aligned}
\end{equation}
\noindent
The small $x$ approximation is now:

\begin{equation*}
P_{xg_2L}(x)=\frac{|a_2|}{\sqrt{2\pi}}{
\frac{\exp\big[-(x+\frac{a_3}{a_3+a_2})^2\frac{(a_3+a_2)^2}{2(\sigma_3^2(1+x)^2)}\big]}
{\sigma_3(1+x)^2}}\,.
\end{equation*}

\noindent
The Cauchy-like term is absent when approximating
the $P_2(z(-1-x)/x)$ as a Dirac $\delta$-function in the
integration of equation~\ref{eq:equation_1} .
The factor $(1+x)$ is retained because it is contained in
the argument of the $\delta$-function. It is essential
to obtain the maximum of $P_{xg_2L}(x)$ in the expected position
$-a_3/(a_3+a_2)$ of its noiseless COG.

\subsection{Probability distribution for $y=\frac{\xi-\eta}{2(\xi+\eta)}$}

Another type of COG$_2$ algorithm is of frequent use, for example
in ref.~\cite{CMS_2014}.
The main difference of this combination of random variables,
from the previous COG$_2$, is a translation
respect to the standard reference system (centered in
the middle of the strip $\#\,2$).
Now, the reference system is centered on the
right border of the strip $\#\,2$.
This COG$_2$ algorithm has the form:
\begin{equation}
    y=x-\frac{1}{2}=\frac{\xi-\eta}{2(\xi+\eta)}\,.
\end{equation}
Even if this is another direct transformation of eq.~\ref{eq:equation_5},
for completeness we report its general form and the case of gaussian PDF.
\begin{equation}
    P_G(y)=P_{xg_2R}(y+\frac{1}{2})=\int_{-\infty}^{+\infty}P_1(\xi)
    P_2(\frac{1-2y}{1+2y}\xi)\frac{|\xi|}{(y+1/2)^2} \mathrm{d}\,\xi
\end{equation}
In the form of $P_G(y)$, we directly use the substitution of
$A\,\mathrm{erf}(A)$ with $|A|$. In any case
$A\,\mathrm{erf}(A)$ is easily obtained from eq.~\ref{eq:equation_5}.

\begin{equation}\label{eq:equation_6c}
    \begin{aligned}
    P_G(y)=&\Big\{\frac{4\big|a_2(1-2y)\sigma_1^2+a_1(1+2y)\sigma_2^2\big|}
    {\sqrt{2\pi}[(1-2y)^2\sigma_1^2+(1+2y)^2\sigma_2^2]^{3/2}}\exp\big[-
    (\frac{a_1-a_2}{2(a_1+a_2)}-y)^2\\
    &\frac{2(a_1+a_2)^2}{(\sigma_1^2(1-2y)^2+(1+2y)^2\sigma_2^2)}\big]\,
    \Big\}+\\
    &\exp[-\frac{a_1^2}{2\sigma_1^2}-\frac{a_2^2}{2\sigma_2^2}]
    \frac{4\sigma_1\sigma_2}{\pi[(1-2y)^2\sigma_1^2+(1+2y)^2\sigma_2^2]}\\  \,.
    \end{aligned}
\end{equation}
With a similar transformation, the PDF for $y=x+1/2$
can be obtained,  here the reference system is centered
in  left border of strip $\#\,2$ with the strip $\#\,3$.
A discussion of the variance of $y$ for small errors is given
in ref.~\cite{samedov}, even if the variance is an ill defined
parameter due to the Cauchy-(Agnesi)-like tails of the PDF. In
this case the results depend from the assumptions introduced.

These PDFs have simple analytical forms, they are defined in reference
systems that depend from the signal in the second strip. Their use, in
maximum likelihood search, could imply complications in the exploration
of the likelihood surface. In fact, if the maximum is outside the two strips of the
PDF, another function must be introduced with a different reference system.

\subsection{Probability distribution for $w=\frac{\xi}{\eta}$ }

As a final use of eq.~\ref{eq:equation_1}, we apply it to obtain the PDF for the
ratio of random variables  $w=\xi/\eta$. Now it is:
\begin{equation}
    x=\frac{\xi}{\xi+\eta}\  \  \
    \  \ w=\xi/\eta \ \  \  \ x=\frac{w}{1+w}\ \ \  \ \ \
    P_{\xi/\eta}(w)=P_{xg_2R}(\frac{w}{1+w})\frac{1}{(1+w)^2}
\end{equation}
The integral expression of $P_{\xi/\eta}(w)$ becomes:
\begin{equation}
    P_{\xi/\eta}(w)=\frac{1}{w^2}\int_{-\infty}^{+\infty}P_1(z)P_2\big(\frac{z}{w}\big)\,|z| \mathrm{d} z
\end{equation}
and transforming eq.~\ref{eq:equation_5} in $w$, as indicated, the  $P_{\xi/\eta}(w)$
for Gaussian PDFs becomes:
\begin{equation}
    \begin{aligned}
    P_{\xi/\eta}(w)=&\Big\{\frac{a_2\sigma_1^2+a_1w\sigma_2^2}{\sqrt{2\pi}(\sigma_1^2+w^2\sigma_2^2)^{3/2}}\exp[-
    \big(\frac{a_1}{a_2}-w\big)^2\frac{a_2^2}{2(\sigma_1^2+w^2\sigma_2^2)}]\,\\
    &\mathrm{erf}[\frac{a_2\sigma_1^2+a_1w\sigma_2^2}
    {\sqrt{2}\sigma_1\sigma_2\sqrt{\sigma_1^2+w^2\sigma_2^2}}]\Big\}+
    \exp\big[-\frac{a_1^2}{2\sigma_1^2}-\frac{a_2^2}{2\sigma_2^2}\big]\frac{\sigma_1\sigma_2}{\pi(\sigma_1^2+w^2\sigma_2^2)}\\
    \end{aligned}
\end{equation}
The last term with $a_1=a_2=0$ coincides with that reported in
ref.~\cite{gnedenko}. Now the  maximum of the first term is
moved to be around $a_1/a_2$.

\section{The PDF of the complete COG$_2$ algorithm}

To obtain the PDF for the COG$_2$ algorithm, we have to define in
detail this algorithm. As previously recalled, we have to
consider the signals of three strips: the strip with the
maximum signal (strip $\#\,2$) and the two lateral (strip $\# 1$ to
the right and strip $\# 3$ to the left). Around the strip $\#\,2$
the strip with the maximum signal is selected between the two strips
$\#\,1$ and $\#\,3$. Due to the smallest number of strips,
this COG$_2$ has a very favorable signal-to-noise ratio.
It is the natural selection for orthogonal incidence on
strip detectors with strip widths near to
the average lateral drift of the primary charges.

\subsection{The definition of the complete COG$_2$ algorithm}

The definition of COG$_2$ algorithm
can be condensed in the following equation (ref.~\cite{landi06}):

\begin{equation}\label{eq:equation_7}
    x_{g2}=\frac{x_1}{x_1+x_2}\theta(x_1-x_3)-\frac{x_3}{x_3+x_2}\theta(x_3-x_1)\,.
\end{equation}
Where $x_1,\,x_2,\,x_3$ are the random signals of the three strips,
and $\theta(z)$ is the Heaviside $\theta$-function ($\theta(x)=0$
for $x\leq 0$ and $\theta(x)=1$ elsewhere). The two $\theta$-functions
select the strip with the highest signal. No condition is
imposed on the strip $\#\,2$, even if for its role of seed strip, it
has some constraints. This choice eliminates inessential complications
and saves the normalization of the PDF.
Our aim is to reproduce
the gap for $x_{g_2}\approx\,0$, typical of the histograms of COG$_2$ algorithm.
This gap is given by the impossibility (or lower probability)
to have $x_{g_2}\approx 0$ if the charge
drift populates one or both the two lateral strips.
The gap grows rapidly with an increase of these two charges.
The noise allows the forbidden values, promoting a lower noiseless signal
to become higher than the other.

The constraints of eq.~\ref{eq:equation_7},
on the three random signal $\{x_1,x_2,x_3\}$, are
inserted in the integral for the PDF of
this COG$_2$: $  P_{x_{g2}}(x) $.
Its integral expression is given by (with
the usual substitution of $x_{g_2}$ as $x$):

\begin{equation}\label{eq:equation_8}
\begin{aligned}
    P_{x_{g2}}(x)=&\int_{-\infty}^{+\infty}\mathrm{d}\,x_1\,
    \mathrm{d}\,x_2\, \mathrm{d}\,x_3  P_1(x_1)P_2(x_2)P_3(x_3)
    \Big[\delta\big(x-\frac{x_1}{x_1+x_2}\big)\theta(x_1-x_3)+\\
    &\delta\big(x+\frac{x_3}{x_3+x_2}\big)\theta(x_3-x_1)\Big] \,.
\end{aligned}
\end{equation}
The normalization of $P_{x_{g2}}(x)$ can be proved with a direct
$x$-integration. The other integrals are executed with the
transformations: $x_1=\xi$, $x_1+x_2=z_1$,  $x_3=\beta$
and $x_3+x_2=z_2$. The jacobian of each couple of transformations
is one, the integrals on  $z_1$ and $z_2$ of the two $\delta$-functions
can be performed with the rule:

\begin{equation}\label{eq:equation_9}
    \int_{-\infty}^{+\infty}\mathrm{d}\,z\, F(z)\delta\big(x\pm\frac{\mu}{z}\big)=
    F\big(\mp\frac{\mu}{x}\big)\frac{|\mu|}{x^2} \,.
\end{equation}
The general form of $P_{x_{g2}}(x)$ for any type of
signal PDF $\{P_1,P_2,P_3\}$ becomes:

\begin{equation}\label{eq:equation_10}
\begin{aligned}
    P_{x_{g2}}(x)=\frac{1}{x^2}\Big[&\int_{-\infty}^{+\infty}\mathrm{d}\xi P_1(\xi)P_2(\xi\frac{1-x}{x})|\xi|\int_{-\infty}^{\xi}\mathrm{d}\beta P_3(\beta)+\\
    &\int_{-\infty}^{+\infty}\mathrm{d}\beta P_3(\beta)P_2(\beta\frac{-1-x}{x})|\beta|
    \int_{-\infty}^{\beta}\mathrm{d}\xi P_1(\xi)\Big] \,.
\end{aligned}
\end{equation}
The gaussian PDFs of eq.~\ref{eq:equation_2}, inserted in
eq.~\ref{eq:equation_10}, allow the explicit expression of the two integrals
on $P_3(\beta)$ and $P_1(\xi)$ with the appropriate
erf-functions. Indicating the remaining integration variable  as $z$,
eq.~\ref{eq:equation_10} becomes:

\begin{equation}\label{eq:equation_11}
\begin{aligned}
& P_{xg_2}(x)=\frac{1}{2\pi\sigma_1\sigma_2 x^2}\\
&\Big(\int_{-\infty}^{+\infty} \mathrm{d}
z \,|z|\Big\{\exp\big[-\frac{(z-a_1)^2}{2\sigma_1^2}-
\frac{(\frac{(1-x)z}{x}-a_2)^2}{2\sigma_2^2}\big]\,\frac{1}{2}
\big[1-\mathrm{erf}(\frac{a_3-z}{\sqrt{2}\sigma_3})\big]+\\
&\frac{\sigma_1}{\sigma_3}\exp\big[-\frac{(z-a_3)^2}{2\sigma_3^2}-
\frac{(\frac{(-1-x)z}{x}-a_2)^2}{2\sigma_2^2}\big]\,\frac{1}{2}
[1-\mathrm{erf}(\frac{a_1-z}{\sqrt{2}\sigma_1})]\Big\}\Big) \,.
\end{aligned}
\end{equation}
The combination of erf-functions and the $|z|$ render impossible
an analytical integration of eq.~\ref{eq:equation_11}. The serial
development of the erf-function and its successive integration term by term is
too cumbersome to be of practical use.  Thus, we
have to explore approximations apt to be useful in maximum likelihood search.

\subsection{Small |x| approximation}

The small $|x|$ approximation is one of the easiest way
to handle eq.~\ref{eq:equation_11}. The function
$P_2(z(1-x)/x)$ can be transformed to
approximate a Dirac $\delta$-function for small $|x|$:

\begin{equation}
\begin{aligned}
    &\frac{\exp\Big[-\big(\frac{1-x}{x}z-a_2\big)^2\frac{1}
    {2\sigma_2^2}\Big]}{\sqrt{2\pi} \sigma_2}=
    \frac{\exp\Big[-\big(\frac{z}{a_2}-\frac{x}{1-x}\big)^2
    \frac{(1-x)^2\,a_2^2}{2x^2\,\sigma_2^2} \Big]}
    {\Big(\frac{\,\sqrt{\,2\,\pi\,}\, \sigma_2\, |\,x\,|\,}{\,a_{\,2}\,|\,(1-x)\,|\,}\Big)}
    \frac{|x\,|}{a_2|(1-x)|}\\
    &\approx \frac{|x\,|}{a_2|(1-x)|}\delta(\zeta-\frac{x}{1-x})
    \ \ \ \ \ \ \ \zeta=\frac{z}{a_2}  \,.
\end{aligned}
\end{equation}
The effective standard deviation of the gaussian is $\sigma_2 |x|/(a_2 |1-x|)$,
this term, for $|x|\rightarrow 0$, allows to identify
the gaussian with a Dirac $\delta$-function. The term $|1-x|$ is useful to obtain
the combination $a_1/(a_1+a_2)$ in the exponent of the Gaussian-like function.
A similar transformation can be applied to $P_2(z(-1-x)/x)$, the integration on $\zeta$
is now immediate and the small $|x|$ probability $P_{xg_2}$ becomes:

\begin{equation}\label{eq:equation_13}
\begin{aligned}
    P_{xg_2}(x)=&\frac{|a_2|}{2\sqrt{2\pi}}\Big\{\frac
    {\exp\big[-(x-\frac{a_1}{a_1+a_2})^2\frac{(a_1+a_2)^2}{2(\sigma_1^2(1-x)^2)}\big]
    \big(1-\mathrm{erf}\big[(\frac{a_3}{a_3+a_2}-x)\frac{a_2+a_3}
    {\sqrt{2}(1-x)\sigma_3}\big]\big)}{\sigma_1(1-x)^2}+\\
    &\frac{\exp\big[-(x+\frac{a_3}{a_3+a_2})^2\frac{(a_3+a_2)^2}{2(\sigma_3^2(1+x)^2)}\big]
    \big(1-\mathrm{erf}\big[(\frac{a_1}{a_1+a_2}+x)\frac{a_1+a_2}{\sqrt{2}(1+x)\sigma_1}\big]\big)}
    {\sigma_3\,(1+x)^2}\Big\} \,.
\end{aligned}
\end{equation}
The term $a_2$ is a positive constant
(the charge of the seed strip) and the absolute value
can be eliminated, but for future developments is
better to remember its presence.
It is easy to recognize in eq.~\ref{eq:equation_13} the two maxima
of eq.~\ref{eq:equation_5} and eq.~\ref{eq:equation_6}, the noiseless
position of the two branch of the COG$_2$ algorithm.
The main difference is due to the two $(1-\mathrm{erf}(-z))/2$-functions,
this type of functions are similar to a continuous
(and derivable) Heaviside-$\theta$ functions. They interpolate
in a very smooth way the two sides of the PDF.
Two different simulated distributions are reported
in refs.~\cite{landi05,landi06} and compared with
eq.~\ref{eq:equation_13}, the first was without Landau fluctuations
and the second contained approximate Landau fluctuations.
At orthogonal incidence, the Landau fluctuation is well
described by the fluctuation of the total released charge.

The approximation of eq.~\ref{eq:equation_13} reproduces, in a
reasonable way, the COG$_2$ PDF
even for non small $x$. In fact, the real useful range of $x$ is $|x|\leq 0.5$,
and the factor that is supposed small is $|x|\sigma_2/a_2$. But, the constant
$a_2$ is connected to seed
of the cluster and it has a high probability to be larger
than few times $\sigma_2$. Its noisy detected part, $x_2$, must assure a
reasonable detection efficiency of the hit. Surely eq.~\ref{eq:equation_13} drops
out at $x=\pm 1$.
In any case better approximations are always useful, given that the probability
$P_{xg_2}(x)$ has to apply to a large set of experimental configurations.
A conceptual incompleteness of eq.~\ref{eq:equation_13} is the lack of the
normalization. The normalization assures a constant probability of the impact
point (eq.~\ref{eq:equation_1a}) but its lack is not a real limitation
for the practical use of eq.~\ref{eq:equation_13}.

\subsection{A better approximation for $P_{xg_2}(x)$}

A more accurate approximation for $P_{xg_2}(x)$ can be obtained
retaining the small $x$ approximation for the two
$\mathrm{erf}$-function of eq.~\ref{eq:equation_11} and
integrating on $z$ the remaining parts. Now the two
integrals have analytical forms, one identical to
eq.~\ref{eq:equation_5} (a part a factor $1/2$) and the other to
eq.~\ref{eq:equation_6}. This approximation saves even
the normalization, obviously within the precision of
a numerical integration of a heavy-tail PDF. We have to
remind that the normalization is the only converging
integral of all our PDFs.

As usual  we substitute $A\mathrm{erf}(A)$
with $|A|$. In any case, the expressions of
the $\mathrm{erf}$-functions are that of eq.~\ref{eq:equation_5} and
eq.~\ref{eq:equation_6}.
Even the Cauchy-like terms are neglected. They
are very small. For example, the seed charge in
some experiments is selected to be around
$6 \sim 7\sigma$, thus the term $\exp(-a_2^2/2\sigma_2^2)$
could be around $\exp(-18)\approx 10^{-8}$.
In some special condition, these terms could be
useful for the outliers hit suppression~\cite{landi05}
that depend from the PDF tails. As previously stated, they assure
the strict positivity of $P_{xg_2}(x)$ even for $a_1=a_2=a_3=0$.
However, we did not insert them in our track reconstructions.

\begin{equation}\label{eq:equation_17}
    \begin{aligned}
    P_{xg_2}(x)=&\Big\{\frac{\Big|a_2(1-x)\sigma_1^2+a_1x\sigma_2^2\Big|}
    {2\sqrt{2\pi}[(1-x)^2\sigma_1^2+x^2\sigma_2^2]^{3/2}}\exp\big[-
    (\frac{a_1}{a_1+a_2}-x)^2\frac{(a_1+a_2)^2}{2(\sigma_1^2(1-x)^2+x^2\sigma_2^2)}\big]\Big\}\\
    &\Big\{1-\mathrm{erf}\big[(\frac{a_3}{a_3+a_2}-x)\frac{a_2+a_3}
    {\sqrt{2}(1-x)\sigma_3}\big]\Big\}+ \  \ a_1\leftrightarrow a_3,
    \ \sigma_1\leftrightarrow\sigma_3, \ \ x\rightarrow -x\,.
    \end{aligned}
\end{equation}
\noindent
An easy simulation can be done to verify equation~\ref{eq:equation_17} and
to illustrate the weak gap present  in a
distribution of simulated $x_{g_2}$  (figure 14) in ref.~\cite{landi05}.
The data are generated
with the function {\tt randn} of MATLAB and with the
equations $x_i=\sigma$\,{\tt randn(1,N)}$+a_i$, inserted in
equation~\ref{eq:equation_7}. Realistic values for $a_i$,
$\sigma_i$ can be obtained from ref.~\cite{landi05} for orthogonal
incidence on the two types of silicon detectors studied there.

\begin{figure} [h!]
\begin{center}
\includegraphics[scale=0.75]{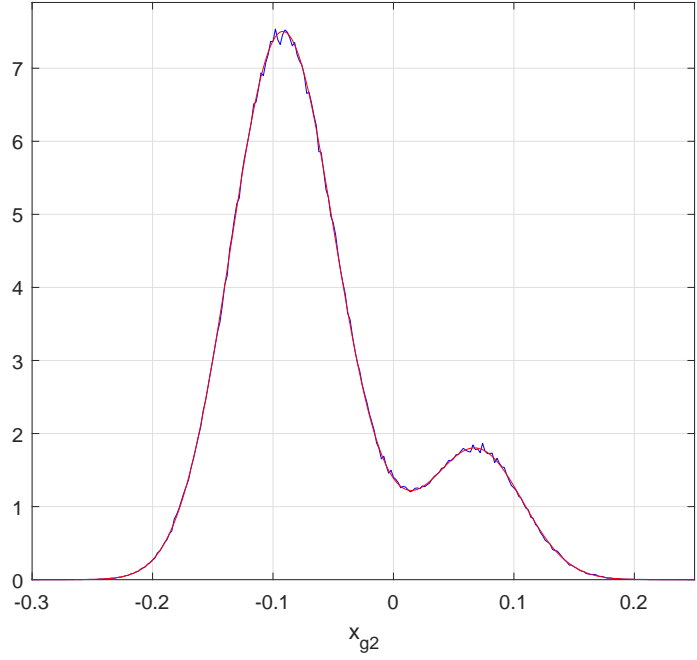}
\caption{\em Empirical PDFs of $x_{g_2}$ (blue line)
compared with equation~\ref{eq:equation_17} (red line) for a
model of silicon detector of ref.~\cite{landi05} for an impact
point $\varepsilon=-0.2$, all the $\sigma=8$ ADC counts, and $a_1=0.01\,E$,
$a_2=0.91\,E$ and $a_3=0.08\,E$ are the charges collected by the three
strips, here $E$, the total charge of the three strips, is 150 ADC counts}
\label{fig:figure_1}
\end{center}
\end{figure}
\noindent
The probability decrease  between the principal and secondary
maximum of figure~\ref{fig:figure_1} originates a similar
reduction in a vertical section of figure~14 of ref.~\cite{landi05}.
The secondary maximum is produced by the noise that promotes the minority
noiseless signal to becomes the greater one. Signal clusters
with lower total charge show larger gaps.

\noindent
Even if equation~\ref{eq:equation_17} represents a better approximation
compared to eq.~\ref{eq:equation_13},  in some extreme cases,
it shows appreciable deviation respect to the numerical
integral of eq.~\ref{eq:equation_11}.
For example, for tracks with large inclination, the combination of parameters
$a_1/(a_1+a_2)$ are very near to $a_3/(a_3+a_2)$ and
slightly lower than $0.5$. In this case the two maximums are widely separated
and the $P_{xg_2}(x)$ of equation~\ref{eq:equation_17} shows discrepancies
compared to the numerical integration
of eq.~\ref{eq:equation_11}. These  discrepancies are absent
in the longer approximation reported in Appendix B.

%
%

\section{Simplified form of the three strip COG}

To test the quality of the functional forms of the $\{a_i(\varepsilon)\}$,
the reconstruction of the three-strip COG (COG$_3$) histograms were extensively
used in ref.~\cite{landi05}, for this, the COG$_3$ PDF was essential.
We will not discuss here the full form of the COG$_3$ PDF with its gaps
at the strip borders as illustrated in ref.~\cite{landi01}. This incomplete PDF
is useful in all the cases when the border gaps are very small (near orthogonal incidence).
\begin{equation}\label{eq:equation_19}
   P_{xg_3}(x)= \int_{-\infty}^{+\infty}\,\mathrm{d} x_1\,
   \mathrm{d} x_2\, \mathrm{d} x_3 P_1(x_1) P_2(x_2) P_3(x_3)
   \delta\big(x-\frac{x_1-x_3}{x_1+x_2+x_3}\big)\,.
\end{equation}
\noindent
Again, the normalization of $P_{xg_3}(x)$ is easily verified. The
substitution of variables $\xi=(x_1-x_3)$,  $z=(x_1+x_2+x_3)$
and $\beta=x_2$ simplifies the Dirac $\delta$-function integration.
The jacobian of the substitution
is $1/2$. Integrating in $\xi$ the Dirac $\delta$-function, the
remaining double integral has the following form:

\begin{equation}\label{eq:equation_20}
    P_{xg_3}(x)=\frac{1}{2}\int_{-\infty}^{+\infty}\,\mathrm{d}z\,
    \mathrm{d}\beta |z| P_1(\frac{z(1+x)-\beta}{2})
    P_2(\beta)\,P_3(\frac{z(1-x)-\beta}{2})\,.
\end{equation}
\noindent
The integration in $\beta$ is a convolution of gaussian PDFs and it gives another
gaussian. Due to the  $|z|$, the integral on $z$
produces the term of the form $A\,\mathrm{erf}(A)$ that, as usual, we
approximate as $|A|$.
Equation~\ref{eq:equation_20} does not contain the explicit term $1/x^2$
of equation~\ref{eq:equation_9}, this is due to the integration in $\xi$,
in any case, the Cauchy-like tails remain.
The introduction of the auxiliary constants $X_3$ and $E_3$ simplifies the form of
$P_{xg_3}(x)$. The Cauchy term, indicated with $P_{xg_3}^C(x)$, is
the first discussed. It has the expression:

\begin{equation}
\begin{aligned}
    X_3=&(a_1-a_3)/(a_1+a_2+a_3) \ \ \ \ \ \ E_3=a_1+a_2+a_3\\
    &  \ \ \ \\
    P_{xg_3}^C(x)=&\exp\big[-\frac{E_3^2\,[\sigma_1^2(X_3-1)^2+
    \sigma_2^2(X_3)^2+\sigma_3^2(X_3+1)^2]}
    {2(\sigma_1^2\sigma_2^2+4\sigma_1^2\sigma_3^2+\sigma_3^2\sigma_2^2)}\big]\\
    &\Big\{\frac{\sqrt{\sigma_1^2\sigma_2^2+4\sigma_1^2\sigma_3^2+\sigma_3^2\sigma_2^2}}
    {\pi[(1-x)^2\sigma_1^2+x^2\sigma_2^2+(1+x)^2\sigma_3^2]}\Big\}\\ \,.
\end{aligned}
\end{equation}
\noindent
The term $P_{xg_3}^C(x)$ for $\sigma_1=\sigma_2=\sigma_3$, as it is often the case,
has the very simple form:
\begin{equation*}
     P_{xg_3}^C(x)=\exp\big[{-\frac{E_3^2\,(X_3^2+2/3)}
     {4\sigma_1^2}}\big]\frac{\sqrt{2/3}}{\pi(x^2+2/3)} \,.
\end{equation*}
\noindent
This term survive even for $E_3=0$ and becomes an exact Cauchy PDF.
The main term $P_{xg_3}(x)$ is:

\begin{equation}\label{eq:equation_21}
\begin{aligned}
    P_{xg_{3}}(x)=&\left\{\exp\Big[-(X_3-x)^2\frac{E_3^2}
    {2[(1-x)^2\sigma_1^2+x^2\sigma_2^2+(1+x)^2\sigma_3^2]}\Big]\right\}\\
    &\frac{\,\,\Big|E_3\,[(1-X_3)(1-x)\sigma_1^2+X_3x\sigma_2^2+(1+X_3)(1+x)
    \sigma_3^2]\Big|\,\,}
    {\sqrt{2\pi}\Big[(1-x)^2\sigma_1^2+x^2\sigma_2^2+(1+x)^2\sigma_3^2\Big]^{3/2}}\,,\\
\end{aligned}
\end{equation}
\noindent
the approximation of $A\,\mathrm{erf}(A)$ as $|A|$ has no observed
differences in our realistic simulations.
In any case, the $\mathrm{erf}$-function is:

\begin{equation}\label{eq:equation_22}
    \mathrm{erf}\left\{\frac{E_3\,\big[(1-X_3)(1-x)\sigma_1^2+X_3x\sigma_2^2+(1+X_3)(1+x)
    \sigma_3^2\big]}{\big[2(\sigma_1^2(1-x)^2+x^2\sigma_2^2+(1+x)^2\sigma_3^2)
    (\sigma_2^2\sigma_3^2+\sigma_1^2(\sigma_2^2+4\sigma_3^2))\big]^{1/2}}\right\}\,.
\end{equation}
\noindent
The upper part of the fraction in the $\mathrm{erf}$-argument in
eq.~\ref{eq:equation_22} coincides with the corresponding term in the absolute value
of eq.~\ref{eq:equation_21}.

A more precise form for the COG$_3$ algorithm should consider the gap  at the
strip borders. This happens when the signal distribution is
larger than two strips, and $x\approx 1/2$ is suppressed in the COG$_3$
(ref.~\cite{landi01} contains other details).
The suppression increases rapidly as the (average) signal distribution
grows beyond the two strip size.
Near to the strip borders, the noise can
increase the signal collected by the nearby strip
that becomes the seed of another three-strip cluster.
In this case, the COG$_3$ algorithm operates with
the triplet of signals $\{x_2,x_1,x_4\}$ where $x_4$
is the signal of the strip  to the right of the strip $\# 1$.
The form of the algorithm becomes:

\begin{equation}\label{eq:equation_23}
    x_{g_3}=\Big(\frac{x_1-x_3}{x_1+x_2+x_3}\Big)\theta(x_2-x_1)+
    \Big(\frac{x_4-x_2}{x_1+x_2+x_4}+1\Big)\theta(x_1-x_2) \,.
\end{equation}
\noindent
The two sides of eq.~\ref{eq:equation_23} are defined
in the identical reference system centered on the strip $\# 2$.
The details of this extension of the COG$_3$ algorithm will be
reported elsewhere.

\section{Conclusions}

This is a first part of a study for COG PDFs, essential tools to go beyond
the methods based on variance minimizations. The long analytical equations,
reported here, are indispensable components to implement the maximum likelihood search.
Even if complex and slow, the maximum likelihood could be able to obtain
results impossible with other methods. For example, the elimination of the
effects introduced by the outliers. Among other beneficial effects, the increase of the
track-parameter resolution could reduce the complexity of the tracker hardware,
requiring less detection layers (or less magnetic field) to obtain the
resolution of the standard
least squares method (or of its equivalent Kalman filter).
These equations were on our desk for a long time, but the huge length of
the standard demonstrations forbade their publications. The method,
illustrated here, allowed manageable demonstrations. The produced
expressions can be handled with the essential help of MATHEMATICA.
Numerical simulations with MATLAB complete the verification of the full
process.

\section{Appendix A}

We report here a synthetic calculation of the PDF for
$\xi/(\xi+\eta)$ along the lines of ref.~\cite{gnedenko}
for the ratio of two random variables. The PDF is obtained
differentiating the cumulative probability distribution for
the random variable $x$. The cumulative distribution is
defined as the probability to have $\xi/(\xi+\eta)\leq x$.
Thus, the product of $P_1(\xi)P_2(\eta)$ must be integrated
on regions of the plane $\eta,\xi$ compatibles with the
defined condition.

We have to select two different procedures, one for
$x\leq 0$ and one for $x>0$. The two lines of equation $\xi+\eta=0$ and
$\xi(1-x)/x=\eta$ are the boundaries of the integration
regions. The first line is fixed and separates the two regions
with different signs of the denominator of $\xi/(\xi+\eta)$.
The other line rotates around the origin as $x$ increases and
it is the second boundary of the integration regions.
It overlaps the line $\xi+\eta=0$ when $x\rightarrow \pm \infty$.
The $\eta$-axis separates the two regions with $x\neq 0$.

\begin{figure} [h!]
\begin{center}
\includegraphics[scale=0.641]{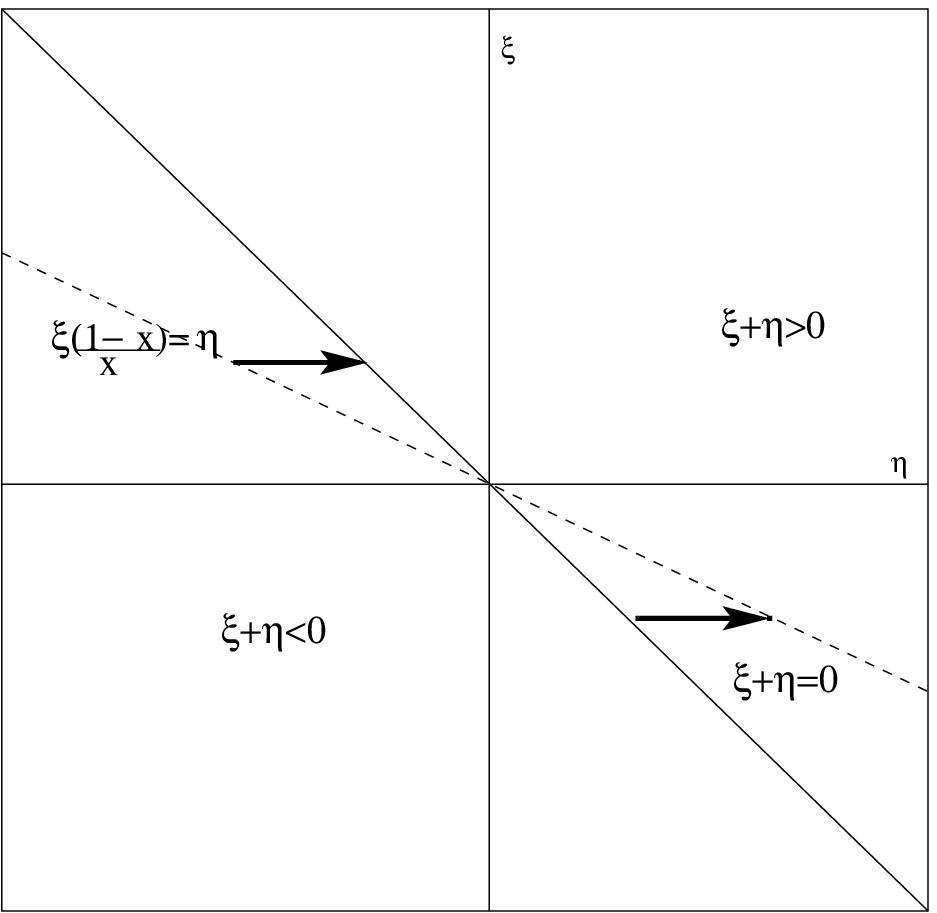}
\includegraphics[scale=0.5]{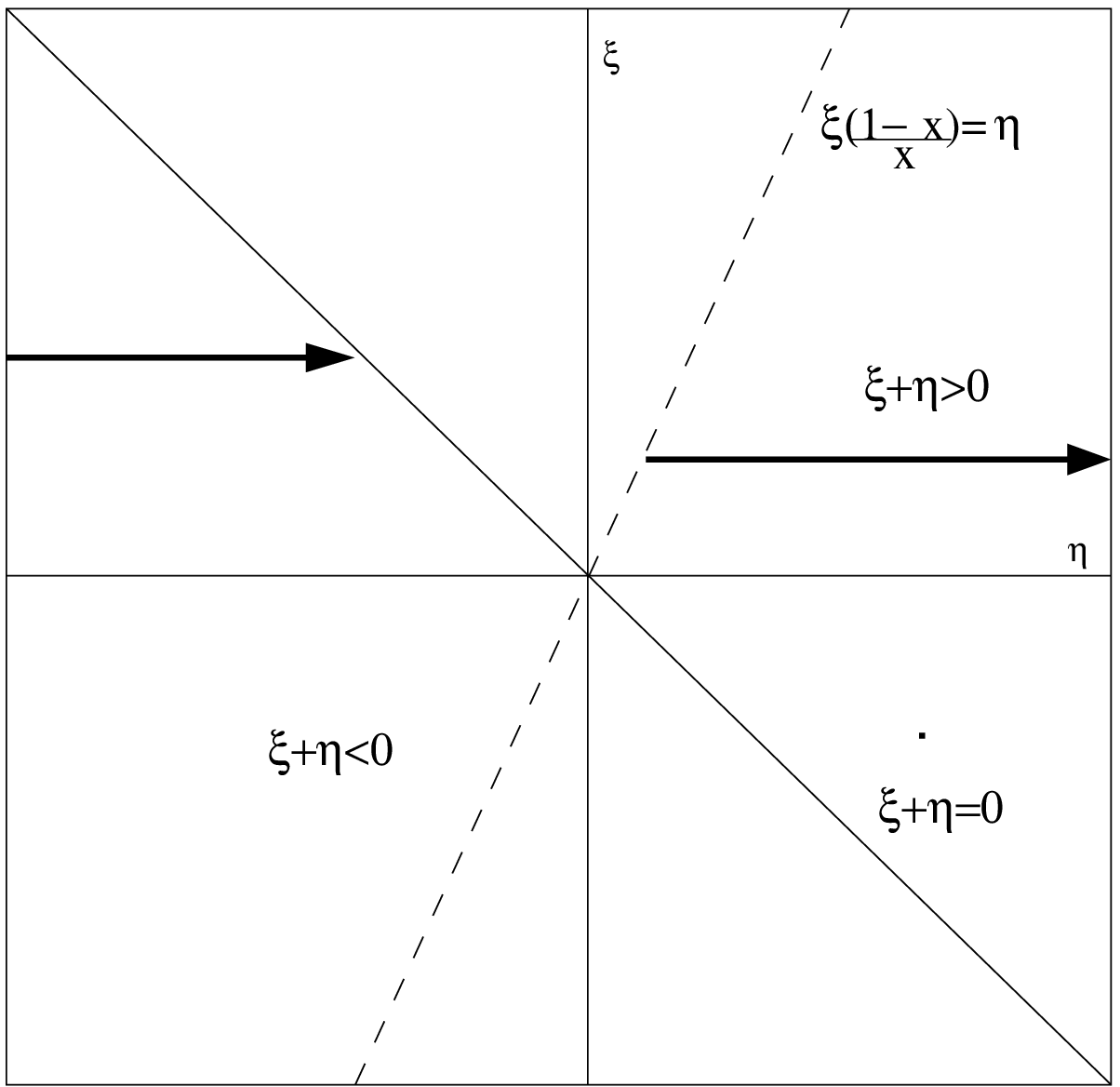}
\caption{\em To the left, the integration regions of the
plane $(\eta,\xi)$ for $x\leq 0$. To the right,
the integration regions for $x>0$. The integration
regions are indicated by thick arrows along
the $\eta$ integrations. The $\xi$ integrations are
not indicated, they are orthogonal the thick arrows to
cover the sector of the plane with the arrows}\label{fig:plot_1}
\end{center}
\end{figure}

\noindent
For $x\leq 0$ we obtain:
\begin{equation}
   F_2^-(x)=\int_{-\infty}^0\,\mathrm{d}\xi P_1(\xi)\int_{-\xi}^{\xi(1-x)/x}
   P_2(\eta)\,\mathrm{d}\eta+\int_0^{\infty}\,\mathrm{d}\xi P_1(\xi)\int_{\xi(1-x)/x}^{-\xi}
   P_2(\eta)\,\mathrm{d}\eta
\end{equation}
\noindent
and for $x>0$ $F_2(x)$ is:
\begin{equation}
\begin{aligned}
F_2^+(x)=&\int_{-\infty}^0\,\mathrm{d}\xi
 P_1(\xi)\int_{-\xi}^{+\infty}P_2(\eta)\,\mathrm{d}\eta+\int_0^{+\infty}\,
 \mathrm{d}\xi\,P_1(\xi)\int_{\xi(1-x)/x}^{+\infty}
P_2(\eta)\,\mathrm{d}\eta+\\
&\int_{-\infty}^0\,\mathrm{d}\xi
P_1(\xi)\int_{-\infty}^{\xi(1-x)/x}P_2(\eta)\,\mathrm{d}\eta+
\int_0^{+\infty}\,\mathrm{d}\xi\,P_1(\xi)\int_{-\infty}^{-\xi}
P_2(\eta)\,\mathrm{d}\eta \,.
\end{aligned}
\end{equation}
It is easy to prove that $F_2^-(x)= 0$ for $x\rightarrow -\infty$ and $F_2^+(x)= 1$
for $x\rightarrow +\infty$.

\noindent
The PDF $P_{xg_2R}(x)$ is given by a differentiation of $F_2^-(x)$ and $F_2^+(x)$ respect to $x$, obtaining:

\begin{equation}
    P_{xg2R}(x)=\frac{\,\mathrm{d} F_2^+(x)}{\,\mathrm{d} x}=
    \frac{1}{x^2}\int_{-\infty}^{+\infty}\,\mathrm{d}\xi\,|\xi|P_1(\xi)
    P_2(\xi\frac{1-x}{x})\,,
\end{equation}
an identical result is obtained differentiating $F_2^-(x)$.
\noindent
The cumulative distribution for the random variable
$-\beta/(\beta+\eta)$ could be obtained with a
similar procedure.

The construction of the cumulative distribution for the
complete COG$_2$ algorithm of equation~\ref{eq:equation_7}
implies the insertion of
another random variable $\beta$. The integration regions
are defined in the space $\xi,\eta,\beta$.
The cumulative distribution is expressed by a large number
of integrals on sectors of the $\xi,\eta,\beta$-space.
The differentiation and the collection of the various terms
reproduces equation~\ref{eq:equation_10}.

\section{Appendix B}

For very inclined tracks, the MIP signal is spread among various
strips and the histograms of COG$_2$ algorithm
show very large gaps around zero. In this case, the approximations
described above show perceptible deviations from the simulated
data and the numerical integrations of eq.~\ref{eq:equation_11}.
In these case a better approximation is useful.
The following approximation shows negligible differences from
the numerical integrations. For its construction, the Fubini theorem
is applied to invert the order of the double integrals
of eq.~\ref{eq:equation_10}, and variable transformations
are selected to have a zero as the lowest limit of an
internal integration region.
In this way the two integrations become independent and
can be executed in any order.
The neglecting of the change of sign introduced by the
absolute values of eq.~\ref{eq:equation_11} allows
to obtain the following analytic result:
\begin{equation}\label{eq:equation_30}
    \begin{aligned}
    &P_{xg2}(x)=\\
     &\frac{1}{2\sqrt{2\pi}}\frac{a_2(1-x)\sigma_1^2+a_1x\sigma_2^2}{[(1-x)^2\sigma_1^2+x^2\sigma_2^2]^{3/2}}
     \,\exp\big[-\big(x-\frac{a_1}{a_1+a_2}\big)^2\frac{(a_1+a_2)^2}{2(\sigma_1^2(1-x)^2+x^2\sigma_2^2)}\big]\\
     &\Big\{1-\mathrm{erf}\big[\frac{(1-x)\big[a_3(1-x)-a_2
     x\big]\sigma_1^2-(a_1-a_3)x^2\sigma_2^2}{\sqrt{2((1-x)^2\sigma_1^2+x^2\sigma_2^2)
     (x^2\sigma_1^2\sigma_2^2+(1-x)^2\sigma_1^2\sigma_3^2+x^2\sigma_2^2\sigma_3^2)}}\big]\Big\}+\\
     & + x\rightarrow -x, \ \ \ a_1\leftrightarrows a_3 \ \ \ \ \sigma_1\leftrightarrows\sigma_3\\
    \end{aligned}
\end{equation}

%
%
\noindent
The approximation does not reproduces the absolute value
(in reality $A\,\mathrm{erf}(A)$) of the previous
equation~\ref{eq:equation_17}. But, for realistic values of the parameters
$\{a_j\}$, it is irrelevant. In any case, it is a trivial
completion if needed, as in~\ref{eq:equation_17}.
Here, the argument of the erf-function is more complete
than that given for small $|x|$ of equation~\ref{eq:equation_17}.
The difference of equation~\ref{eq:equation_30} with a numerical
integration is negligible in many significant cases.

To complete the approximation, we report the Cauchy-like
terms (even  if of scarce
relevance). They are exactly given by MATHEMATICA
because they are the first terms of a by
part integration of equation~\ref{eq:equation_10}.
The Cauchy tails are evident and the factor $x$ in the
numerator compensates the $\sqrt{x^2}$
in the denominator. The expression of $P_{xg2}^{cauchy}(x)$ is:
\begin{equation}
    \begin{aligned}
    &P_{xg2}^{cauchy}(x)=\\
    & \Big\{ \exp\Big[-\frac{(a_3(1-x)-a_2 x)^2\sigma_1^2+(a_1(1-x)-a_2 x)^2\sigma_3^2
    +(a_1-a_3)^2x^2\sigma_2^2}{2\big(x^2\sigma_2^2\sigma_3^2+(1-x)^2\sigma_1^2\sigma_3^2+
    x^2\sigma_2^2\sigma_1^2\big)}\Big]\\
    &\frac{x\,\, \sigma_1^2\,\sigma_2^2}{2\pi\big((1-x)^2\sigma_1^2+x^2\sigma_2^2\big)
    \sqrt{(1-x)^2\sigma_3^2\sigma_1^2+x^2\sigma_2^2(\sigma_1^2+\sigma_3^2)}}\\
     &\mathrm{erf}\Big[\frac{(1-x)a_2\sigma_1^2\sigma_3^2+
    \sigma_2^2(\sigma_1^2a_3+\sigma_3^2 a_1)x}
    {\sqrt{2}\sigma_1\sigma_2\sigma_3\sqrt{ \sigma_2^2
    \sigma_3^2 x^2 +\sigma_1^2(\sigma_3^2(1-x)^2+\sigma_2^2 x^2)}}\Big]\Big\}+\\
    &\exp\big[-\frac{a_1^2}{2\sigma_1^2}-\frac{a_2^2}{2\sigma_2^2}\big]
    \frac{\,\,\big[1-\mathrm{erf}(a_3/\sqrt{2}\sigma_3)\big]\,\sigma_1\sigma_2\,\,}
    {2\pi[x^2\sigma_2^2+(1-x)^2\sigma_1^2] }+\\
    & + x\rightarrow -x, \ \ \ a_1\leftrightarrows a_3 \ \ \ \ \sigma_1\leftrightarrows\sigma_3\,.\\
    \end{aligned}
\end{equation}
These terms are in general a very small fraction of the main terms
(around $10^{-5}$), but become of the order of $10^{-1}$ for very
inclined tracks. In any case they completes the PDF for the COG$_2$ algorithm.
The exponential term has maxima around for $x=a_3/(a_3+a_2)$
(due to the term with $\big(a_1(1-x)+x a_2\big)$) and
$x=a_1/(a_1+a_3)$  (due to the term with $\big(a_3(1-x)+x a_2\big)$),
these two maxima are very near with large overlaps.

%
%



\begin{thebibliography}{99}

\bibitem{landi05} Landi G.;  Landi G. E. {\em Improvement of track reconstruction with well tuned
probability distributions}  {\em JINST} 9 2014 P10006. {\tt arXiv:1404.1968[physics.ins-det]}
https://arxiv.org/abs/1404.1968

\bibitem{landi06} Landi, G.;  Landi G. E.   {\em Optimizing momentum resolution with a new fitting
method for silicon-strip detectors}  { INSTRUMENTS} {\bf 2018}, 2, 22 

\bibitem{landi07} Landi G.; Landi G. E. {\em Beyond the $\sqrt{N}$-limit of the least
 squares resolution and the lucky-model} {\tt arXiv:1808.06708[physics.ins-det]}
 https://arxiv.org/abs/1808.06708.

\bibitem{landi03} G. Landi, {\em Problems of position reconstruction in silicon
microstrip detectors} Nucl. Instr. and Meth. {\bf A 554} (2005) 226.


\bibitem{landi08} Landi G.; Landi G. E. {\em The Cramer-Rao inequality to go beyomd the
$\sqrt{\mathrm{N}}$-limit of the standard least-squares method in track fitting}
{\tt arXiv:1910.14494 [physics.ins-det]} https://arxiv.org/abs/1910.14494.

\bibitem{landi09} Landi G.; Landi G. E. {\em Proofs of non-optimality of the standard
least-squares method for track reconstructions} {\tt arXiv:2003.10021 [math.ST]}

%
\bibitem{landi01} G. Landi, {\em The center of gravity as an algorithm for position measurements}
Nucl. Instr. and Meth. {\bf A 485} (2002) 698 {\tt arXiv:1908.04447 [physics.ins-det]} https://arxiv.org/abs/1910.04447.


\bibitem{MATHEMATICA} MATHEMATICA 6 Wolfram Inc. Champaign IL, USA

\bibitem{matlab} MATLAB 8 The MathWork Inc. Natic, MA, USA


\bibitem{hartmann} F. Hartmann,  {\em Silicon tracking detectors in high-energy physics}
%
{Nucl. Instrum. and Meth.} {\bf A 666} (2012) 25


\bibitem{CMS_13} The CMS Collaboration, {\em The performance of the muon detector in proton-proton
                       collision at $\sqrt{s}=7$ TeV at LHC} JINST 8 (2013) P11002 $\ $ 
                       {\tt arXiv:1306.6905 [physics.ins-det]}

\bibitem{gnedenko} B. V. Gnedenko "The Theory of Probability and Elements of Statistics"
                    (AMS Chelsea Publishing -Providence Rhode Island )

\bibitem{CMS_2014} The CMS Collaboration, {\em Description and Performance of track and
                      primary vertex reconstruction with the CMS tracker.} 2014 JINST 9 P10009
                      {\tt arXiv:1405.6569 [physics.ins-det]}

\bibitem{samedov} V.V. Samedov {\em Inaccuracy of coordinate determined by several
                            detectors' signals} 2012 JINST 7 C06002








\end{thebibliography}
\end{document}